\documentclass[12pt,letterpaper]{article}
\usepackage{latexsym}
\usepackage{amsfonts}
\usepackage{amsmath}
\usepackage{amssymb}
\usepackage[dvips]{color}
\usepackage[dvips]{graphicx}
\usepackage{epsfig}
\usepackage{pstricks}  %for Ubuntu
\usepackage{hyperref}

\newcommand{\be}{\begin{equation}}
\newcommand{\ee}{\end{equation}}
\newcommand{\ba}{\begin{eqnarray}}
\newcommand{\ea}{\end{eqnarray}}

\def\ie{\emph{i.e.}\,}
\def\munu{\mu \nu}
\def\albe{\alpha \beta}

\begin{document}

\begin{center}
{\Large\bf From isospin generators to BRST quantization 
of higher spin massless fields}
\end{center}
\bigskip

\centerline{{\large J. Alfaro}$^{\,a,}$\,%
\footnote{\texttt{E-mail: jalfaro@puc.cl}},
{\large M. Cambiaso}$^{\, b,\,c,}$\,%
\footnote{\texttt{E-mail: mauro.cambiaso@nucleares.unam.mx}}}
\smallskip
\centerline{$^a$
{\it Facultad de F\'{\i}sica, Pontificia 
Universidad Cat\'olica de Chile,
}}
\centerline{\it Vicu\~{n}a Mackenna 4860 Macul, Santiago de Chile,
Casilla 306, Chile }
\smallskip
\smallskip
\centerline{$^b$ {\it Instituto de Ciencias Nucleares,}} 
\centerline{\it Universidad Nacional Aut\'onoma de M\'exico, %
A. Postal 70-543, M\'exico D.F., M\'exico}
\smallskip
\smallskip
\centerline{$^c$ {\it Departamento de Ciencias F\'isicas, Facultad de Ingenier\'ia,}}
\centerline{\it Universidad Andr\'es Bello, Av. Rep\'ublica 252, Santiago, Chile}
\vskip .5cm

% \title{Gauge Invariant Theory for Arbitrary Symmetry Tensor Fields}
% 
% \author{Jorge Alfaro$^1$} 
% \email{jalfaro@uc.cl}
% \author{Mauro Cambiaso$^{2,3}$}
% \email{mauro.cambiaso@nucleares.unam.mx}
% 
% \affiliation{$^1$Facultad de F\'{\i}sica, Pontificia Universidad Cat\'olica de Chile, \\
% Casilla 306, Santiago 22, Chile}
% 
% \affiliation{$^2$Instituto de Ciencias Nucleares, Universidad Nacional Aut\'onoma de M\'exico,\\
% A. Postal 70-543, M\'exico D.F., M\'exico,}
% \affiliation{$^3$Departamento de Ciencias F\'isicas, Facultad de Ingenier\'ia, Universidad Andr\'es Bello, 
% Av. Rep\'ublica 252, Santiago, Chile}
% 

\begin{abstract}

Motivated
by construction of isospin generators %Goergi Lie Algebras in part phys... 2 Ed. pp 83, 1st Ed. Exc V.B
in particle physics (built form 
the SU(2) algebra), we find an equivalence
between the algebra of these generators and those of the Virasoro algebra. The form
of the starting generators is fixed and in order to obtain a full equivalence
we introduce a new matrix product. The Cartan structure of the starting algebra
is reproduced for the Virasoro-like case and a natural BRST quantization as in String
Field Theory is 
``induced'' in the Fock space of the creation/annihilation operators.
Following this procedure, we find a rather trivial Lie algebra 
form which we obtain a gauge theory
of an infinity on non-interacting massless particles of arbitrary integer 
spin and symmetry. 
Among others we find the free Maxwell field, the free (linearized) gravitational field and also the axion field  with their 
appropriate gauge transformations. 

\end{abstract}

\section{Introduction}
Isospin is not only important in  nuclear physics for 
the description of the interactions between protons $p$ and 
neutrons $n$ in the nucleii and other particles participating 
in strong interactions. Most importantly, the idea of 
isospin invariance led Yang and Mills  to the formulation 
of gauge theories \cite{Yang:1954ek}.  

The idea of isospin,
originally due to Heisenberg \cite{Heisenberg:1932dw}, is that in nuclear physics, 
if one forgets about the electromagnetic 
and the weak interactions, the strong interactions are (almost) 
insensitive to the change of protons by neutrons and vice-versa. 
Hence, in this context, there is an (almost) exact symmetry 
and therefore the Hamiltonian of strong interactions $H_S$ 
has (almost) degenerate states with the same energy %
\footnote{ If state $B$ is obtained from state $A$ by exchanging 
$p\leftrightarrow n$, then these states $A,B$ correspond to 
two (almost) degenerate eigenstates of $H_S$ with (almost) 
the same energy.}%
. The failure to be an exact symmetry being 
due to the slight mass difference between the proton and the 
neutron, $m_p = 938,3$ MeV,  $m_n = 939,6$ MeV.
Therefore protons and neutrons could be thought of as two 
states of a single ``particle'', the nucleon doublet $N$,
\begin{equation}
 N = \binom{p}{n}\,.
\end{equation}
As we will see in section \ref{sec:Isospin}, one can build
operators that realize the aforementioned symmetry under
$p \leftrightarrow n$ exchange as:
\be \label{eqn:iso1}
T_a = \frac{1}{2}\sum_{i,j} \sum_{\alpha} a^\dag_{i,\alpha} (\sigma_a)_{ij} a_{j,\alpha}\,,
\ee
where $a^\dag_{i,\alpha} (a_{i,\alpha})$ are creation/annihilation
operators for nucleons of the species $i$ in the sate $\alpha$, and
$(\sigma_a)_{ij}$ are the components of the $a$-th Pauli matrix.

In the present work, however, we do not focus
on isospin generators \textit{per se}, rather, we note that the 
bilineality in creation/annihilation 
operators bears close ressemblance with the Virasoro generators 
of string theory. 

On the other hand, the BRST quantization method has proven to
be very useful in gauge theories (as constrained Hamiltonian 
systems in general) and also in string theory. Therefore
we seek to find a quantum theory describing particles' excitations
from a BRST symmetry based on ``generalized isospin'' operators similar
to those in (\ref{eqn:iso1}) which 
we will define below.
Since the idea of generalized operators is not restrictive
to describe spin-1/2 systems we expect our theory to describe
higher spin (HS) particles as well (this is clear form the
group theoretic point of view in which there is no reason
to consider only spins $\leq$ 2 as candidates for elementary
fields -as Nature seems to suggest- since the Lorentz group
admits representations of arbitrary integer of half-integer spins). 
Since this is a gauge
theory we will be dealing with the massless fields. 

The interest in HS particles dates back to the work of Fierz and Pauli
\cite{Fierz:1939ix} for massive case. This was in the context
of finding a covariant description of particles carrying arbitrary
representations of the Lorentz group. The study of the massless
case was considered in \cite{Fronsdal:1978rb,Fang:1978wz,Curtright:1980yk}.
In \cite{Ouvry:1986dv} free HS gauge fields were studied with its 
underlying BRST structure and in \cite{Koh:1986vg} the interacting
case was obtained. Then the above was generalized to mixed symmetry 
massless fields in 
\cite{Labastida:1986gy,Labastida:1986ft,Labastida:1987kw}.

The importance of HS gauge fields is not merely academic.
One of the most appealing reasons concerns string theory.
The low energy limit 
of string theory is relatively well understood, and for example,
there are several field theoretical descriptions of it as effective
theories. On the contrary, its high energy limit is not clear yet
and neither is there a field theoretical description in that 
limit. However, HS gauge field theory may provide such a description
since there is evidence supporting a relation between
HS gauge fields and the high energy limit of string field 
theory \cite{Siegel:1985tw,Witten:1985cc,Jensen:1996dc,Francia:2002aa,%
Lindstrom:2003mg,Bonelli:2003kh,Sagnotti:2003qa}.
Further developments on the BRST formulation of the problem
we made on \cite{Pashnev:1997rm,Pashnev:1998ti,Pashnev:1998sh}
and after many efforts, an appropriate description
of the interactions of HS massless gauge field was derived 
\cite{Buchbinder:2001bs,Fotopoulos:2007nm,Fotopoulos:2007yq,Bengtsson:2006pw}.
For a recent review see \cite{Fotopoulos:2008ka} and references therein.

The structure of the paper is as follows: In section \ref{sec:Isospin} 
we review the construction of isospin operators. 
The basics of the definition of the BRST
charge and the BRST quantization method is given in 
section \ref{sec:BRST}. Next in 
section \ref{sec:alternative BRST} 
we find an algebra built as ``isospin operators'' but with 
creation/annihilation operators satisfying  commutation relations
as in string theory. For the algebra thus obtained to
mimic the Cartan structure of the Virasoro algebra 
we must implement a new product between the matrix representatives
of the algebra. We also show how do the matrix representations 
look like with the newly introduced product for the SU(2) case. Next we focus
on the simplest closed subalgebra of the Heisenberg
algebra
% {Q_q1,\dots ,Q_n;P_1,\dots,P_n,C} / {Q_i,Q_j} = {P_i,P_j} = {C,C} = {Q_i,C} = {P_i,C} = 0 and {Q_i,P_j} = i\delta_{ij}C.
from which we construct the corresponding BRST charge to initiate our
quantization programme in section \ref{sec:Massless spin1,2 fields}, finding 
the correct equations of motion describing free photons and gravitons.
In section \ref{sec:Mixed 2nd rank} we extend our analysis to include mixed symmetry $2^{nd}$ rank
tensor fields and among the new states we find the axion of string theory. 
In section \ref{sec:higher rank} we
extend our algebra to the Heisenberg algebra thus allowing for higher spin gauge fields of 
arbitrary (integer) spin. We show how our results can be properly understood
as a limiting case of the BRST quantization in string theory and connect with previous 
works on higher spin gauge fields. In section \ref{sec:Concl} we conclude by summarizing our
results and some further comments.

\section{Generalized Isospin operators} \label{sec:Isospin}
The following analysis is mostly taken from \cite{geo} and presented
here for completeness and in order to fix the notation.
As mentioned above the idea of isospin was based on forgetting
about the electromagnetic interaction for the study of nuclear
forces, therefore, interactions must be charge independent.
Recalling the usual spin-1/2 case where the spin operator
acts on the two $J_3$ components of a spin-1/2 representation, 
in this case, charge independence would be granted if 
there existed a conserved sort of 
``spin'' operator acting on the $\binom{p}{n}$ doublet.

Thinking of protons (neutrons) in state $\alpha$ as being 
created from the vacuum $|0\rangle$ by the 
operators $a^\dag_{N,1/2,\alpha}$ ($a^\dag_{N,-1/2,\alpha}$) 
respectively, and annihilated by 
$a_{N,1/2,\alpha}$ ($a_{N,-1/2,\alpha}$) 
we can describe the states of a single proton $p$ in state 
$\alpha$ or and r-nucleon state
with the nucleon $m_i$  in the state $\alpha_i$ as:
\begin{eqnarray}
 |p, \alpha \rangle = a^\dag_{N,1/2,\alpha} |0\rangle \qquad \mathrm{and,}\\
 %|r-nucleons, m_1,\alpha_1;m_2,\alpha_2; \dots m_2,\alpha_r \rangle&=&%
 a^\dag_{N,m_1,\alpha_1}a^\dag_{N,m_2,\alpha_2} \dots a^\dag_{N,m_r,\alpha_r}|0\rangle\,, 
\end{eqnarray}
respectively. The label 
$m_i$ telling whether the nucleon is a proton $N = 1/2$ or a 
neutron $N=-1/2$. The subscript $N$ is to remind that, so far, 
these are creation/annihilation operators acting on nucleons $N$. 
Since protons and neutrons are fermions  the creation/annihilation
operators satisfy anticommutation relations:
\begin{eqnarray}
 \{a_{N,a,\alpha}, a^\dag_{N,b,\beta}\} = \delta_{ab} \delta_{\alpha \beta}\,,\\
 \{a^\dag_{N,a,\alpha}, a^\dag_{N,b,\beta}\} = \{a_{N,a,\alpha}, a_{N,b,\beta}\} = 0 \,.
\end{eqnarray}
Isospin symmetry had to do with $p$, $n$ exchange. This can be realized
by considering the following operators.
\ba
T^+ &=& \frac{1}{\sqrt{2}} \sum_\alpha a^\dag_{N,1/2,\alpha} a_{N,-1/2,\alpha} \, \\
(T^+)^\dag = T^- &=& \frac{1}{\sqrt{2}} \sum_\alpha a^\dag_{N,-1/2,\alpha} a_{N,1/2,\alpha}\,.
\ea
One can verify that the first operator when acting on a general
state of $n$ protons in states $\alpha_1,\dots \alpha_n$ and $m$ neutrons
in states $\beta_1, \dots ,\beta_m$ produces a state of $n$ protons in the
same states as before and if the one neutron in the state, say, $\beta_i$
is in the state $\alpha$, then this neutron is exchanged by a proton in
that same state. The other neutrons are left the same. 
If there are no neutrons in state $\alpha$ the action of
$T^+$ on the general state yields 0. Similarly by considering the action 
of $T^-$ on the general state we can see that it exchanges a proton
by a neutron in the state $\alpha$.
Defining $T_3 \equiv [T^+,T^-]$ it can be checked that:
\ba
T_3 &=& \frac{1}{2} \sum_\alpha (a^\dag_{N,1/2,\alpha} a_{N,1/2,\alpha} - a^\dag_{N,-1/2,\alpha} a_{N,-1/2,\alpha})\,, \\
\left[T^+,T^-\right] &=& T_3 \, , \\
\left[T_3,T^\pm\right] &=& \pm\, T^\pm \,,
\ea
\ie, these operators satisfy the angular momentum algebra as 
was to be expected and they are called the \textbf{isospin generators}.

The $T_3$ operator can be rewritten as (sum over repeated Latin indices understood)

\be
T_3 = \frac{1}{2} \sum_\alpha a^\dag_{i, \alpha} (\sigma_3)_{ij} a_{j, \alpha}\,,
\ee
where $(\sigma_3)_{ij}$ are the components of the third Pauli matrix. In 
general, 
\be
T_a = \frac{1}{2}\sum_\alpha a^\dag_{i, \alpha} (\sigma_a)_{ij} a_{j, \alpha}\,.
\ee
Since not only protons and nucleons are involved in strong interactions,
one can generalize the above construction to the other particles
involved in strong interactions. Thus we can write 
\textbf{generalized isospin generators} as:
\be
\widehat T_a = \frac{1}{\sqrt{2}}\sum a^\dag_{x,i,\alpha} [T^{I_x}_a]_{ij} a_{x,j,\alpha}\,,
\ee
where: the sum is over all particles $x$ in the state $\alpha$; $I_x$ 
is the isospin of particle $x$; $i,j$ take values on the $T_3$ 
values for the particles $x$ and $a^\dag_{x,i,\alpha} (a_{x,i,\alpha})$ are
creator/annihilator operators for the particles of type-$x$ and satisfy
commutation or anticommutation relations whether the particle $x$ is bosonic
or fermionic.

It is a simple task to check that the generalized operators 
$\widehat T_a$ thus built
satisfy the same ``isospin'' algebra as the isospin
operators $T_a$  provided
the creation/annihilation operators satisfy the usual 
commutation/anticommutation relations:
\begin{eqnarray}
[a_{x,i,\alpha}, a^\dag_{y,j,\beta}]_\pm &=& \delta_{xy} \delta_{ij} \delta_{\albe}\,,\\
\left[a^\dag_{x,i,\alpha}, a^\dag_{y,j,\beta}\right]_\pm &=& \left[a_{x,i,\alpha}, a_{y,j,\beta}\right]_\pm = 0.
\end{eqnarray}
In the above, when both $x$ and $y$ are fermions, the $+$ subscript rules and
the anticommutator is used. Otherwise, the $-$ does and the commutator
is considered.

The generalized isospin generators are very similar in form to
the generators of the Virasoro algebra written in terms
of oscillators, namely:
\begin{displaymath}
L_{m} = \left\{ \begin{array}{ll}
\frac{1}{2}  p^{2} + \frac{1}{2} \sum_{1}^{\infty} a^{\mu}_{-n}n~a^{\mu}_{n} \qquad & \textrm{for $m =0$,}
\\
\\
\frac{1}{2} \sum_{-\infty}^{\infty} a^{\mu}_{m-n}~ \sqrt{ |(m-n)n| }~a^{\mu}_{n} & \textrm{for $m \neq 0$.}
\end{array} \right.
\end{displaymath}
Where as is customary in string theory, $a^\mu_{-m} = a^{\dag \mu}_m$. In the following
we will  start form a very simple Lie algebra,
whose generators will play the role of the spin generators above. From them we will 
build Virasoro-like ``generalized isospin'' generators  and pursue  
a BRST quantization programme. Therefore in the next section we will 
review the basics of BRST symmetry.

\section{The BRST charge} \label{sec:BRST}
The BRST charge, $\mathcal{\widehat{Q}}_{BRST}$, is a nilpotent operator that acts upon the states
provided by the theory. Due to its nilpotency, this operator generates a gauge
symmetry of the states, namely, if
$|\phi' \rangle = |\phi \rangle + \mathcal{\widehat{Q}}|\lambda \rangle $ then
$\mathcal{\widehat{Q}}~|\phi\rangle = \mathcal{\widehat{Q}}~|\phi'\rangle$
and since physical states may not depend on the gauge
chosen, the physical states must be invariant under the action of the operator.

Let us remind the definition of the BRST charge and the meaning of
invariance of the physical states under the gauge transformation generated
by $\mathcal{\widehat{Q}}$ in some detail so as to demonstrate
the previous statements.
The BRST charge ($\mathcal{\widehat{Q}}$) is defined in the following way:
\begin{equation} \label{q1}
\mathcal{\widehat{Q}} \equiv c^iX_i - 1/2~ f^k_{~ij}~c^ic^jb_k~, \end{equation}
where $X_i$ are the generators of a Lie algebra with $f^k_{~ij}$ as structure
constants, the variables $c^i$ y $b_j$ are \emph{Grassmannian}
fermionic fields satisfying the following
anticommutation relation
$$
\{c^i,b_j\} = \delta^i_{~j}~.
$$
The nilpotency of the charge ($\mathcal{\widehat{Q}}^2=0$) is easily demonstrated from the
property above, from the algebra defining the generators
$[X_i,X_j]=f^k_{~ij}X_k$ and from the Jacobi
identity satisfied by the structure constants.
The ghost number operator is defined as $U \equiv \sum_i c^ib_i$. A certain
state $|\phi\rangle$ with ghost number $n$ will be BRST invariant if
$\mathcal{\widehat{Q}}~|\phi\rangle = 0$. Now
all there is to do for finding physical states is to give a meaning to the ghost
number operator and search for states that are BRST invariant with a given ghost
number. In principle one would expect physical states not to contain any ghost
or antighost excitations at all so one would look for states with ghost number
zero, but analyzing the ghost number operator, specifically the normal
ordering problem of the fermionic fields, shows that there exists physical
states with ghost number different from zero, as a result of the ordering
of $c$'s and $b$'s but still with no ghost excitations.

Now suppose the state $|\phi\rangle$ has ghost number zero. This means that $|\phi\rangle$ must
be annihilated by the $b_k$'s so
\begin{equation} \label{q2}
\mathcal{\widehat{Q}}~ |\phi \rangle = \sum_i c^i X_i |\phi \rangle ~.\end{equation}
But from the anticommutation relations of the fermionic fields, if $|\phi\rangle$ is
annihilated by the $b_k$'s it cannot be annihilated by the $c^j$'s. Thus if the state
$|\phi\rangle$ is physical (BRST invariant) then it must be invariant under the action
of the generators $X_i$.
\begin{equation} \label{q3}
X_i |\phi\rangle = 0,  \qquad i = 1, \dots ,n. \end{equation}
In string theory, Virasoro algebra is not just a coincidence. In fact it is
intimately connected with basic properties of the theory, such as conformal
invariance (Virasoro generators are the generators of conformal transformations
of the intrinsic coordinates in the world-sheet of strings) and to
the solutions to the Euler-Lagrange equations derived from the Polyakov
action. This algebra, however, is infinite-dimensional and the nilpotency
of the BRST charge, constructed from it, is a bit more subtle due to 
anomalies and other problems related to the normal ordering
of the fermionic fields, all of which are solved if certain conditions
are met. These problems will not concern us here, since, as we will see,
we will circumvent them by restricting to a much simpler case.

\section{BRST quantization with a Lie algebra different from Virasoro algebra} \label{sec:alternative BRST}

Virasoro algebra is certainly very important because without it quantization of string
theories would be a difficult task. Our question is then the following.
Is there something even more fundamental for physical theories in Virasoro
algebra than in strings themselves?
To answer this we focused on the fact that the generators of the Virasoro algebra can
be classified \emph{\`a la} Cartan if one forgets about the central
charge %
\footnote{ The Virasoro algebra is
$[L_m,L_n]=(m-n)L_{m+n}+\frac{c}{12}(m^3-m)\delta_{m+n}$. The second term
on the right side arises from the normal-ordering ambiguities of the
operators that define the Virasoro generators at the moment of quantizing the
commutators of the classical generators.
Then our remark is perfectly true for $L_{-1}$, $L_0$ and $L_{1}$, which
generates a closed subalgebra, without anomaly. Furthermore, for the anomaly to be cancelled
in the general case, a particular value of the space-time dimension $D$ must be taken
since $c = c(D)$. In the case of the bosonic string, the anomaly cancellation condition
demands $D=26$. As can be seen, in our case, our results will be valid in for any $D$, since
for the closed subalgebra chosen, the central charge vanishes.}.
Particularly one generator $L_0$ that
commutes with all the resting generators and that can be identified with the Hamiltonian of the
theory and an infinity of raising and lowering operators
$\{L_{\pm m}\}^{\infty}_{m=1}$.
Our goal then is to seek for Lie algebras whose generators can: be identified with
the Virasoro generators, that exhibit the Cartan structure and that we can
construct a reasonable BRST charge from them.

\subsection{Construction of the Lie Algebra}
Let us define creation/annihilation operators $a^{\nu}_n$ that are related to
the creation/annihilation operators of fields' excitations in string theory
$\alpha^{\nu}_n$ as in (\ref{alpha1}) and (\ref{alpha2}).
\begin{equation}  \label{alpha1}
\alpha^{\mu}_n  =  \sqrt{ |n| }~a^{\mu \dagger}_n \qquad \textrm{si $n<0$} \quad \textrm{and}\end{equation}
\begin{equation}  \label{alpha2}
\alpha^{\mu}_n  =  \sqrt{ n }~a^{\mu}_n \qquad \textrm{si $n>0$}~. \end{equation}

These operators, however are not exactly the same as the usual $a^{\nu}_n$ of string theory.
Their Poisson bracket is defined as:
\begin{equation} \label{ext22}
\{a^{\mu}_n ,a^{\nu}_m\}_{P.B}  = i \sigma(n) \delta_{m+n} \eta^{\mu \nu} ~,\end{equation}
where: $a^{\mu}_0 = p^{\mu}$, $\sigma(n)$ is the sign of $n$ such that
$\sigma(0) = 0$ and of course $a^{\nu}_n$ are creation (annihilation) operators
if $n<0 (>0)$.
Let's define operators $\widehat{T}^i$  as follows:
\begin{equation} \label{ext23}
\widehat{T}^i \equiv \frac{1}{\sqrt{2}}\sum_{\mu} \sum_{n,m \neq 0} a^{\mu}_n ~T^{i}_{nm}~ a_{\mu m} \qquad \qquad i \neq 0.\end{equation}

One can prove that if the matrices  $T^i_{nm}$ satisfy the
algebra $[T^i,T^j]= f^{ij}_kT^k$ then the operators defined in (\ref{ext23})
satisfy an algebra with the same structure constants as the matrices
$T^i_{nm}$, provided the creation/annihilation operators be the usual
ones (and not those in (\ref{ext22})) and that the matrix product is also the
usual one, concerning positive components for the matrices.
Let's compute the Poisson bracket between any two of our generators.
\begin{eqnarray} \label{ext24}
\{\widehat{T}^i,\widehat{T}^j\}_{P.B} & = & \frac{i}{2} \sum_{\mu, \nu} \sum_{m,n,o,p}
T^i _{n,m}~ T^j_{o,p} ~~\{a^{\mu}_n ~ a_{\mu m} , a^{\nu}_o ~ a_{\nu p} \}_{P.B} {} \nonumber\\
& = & {} \frac{i}{2}[a^{\mu}_p~a_{\mu m} ~\sigma(n) ~T^i _{n,m}~ T^j_{-n,p} + a^{\mu}_o~a_{\mu m} ~\sigma(n) ~T^i _{n,m}~ T^j_{o,-n} + {} \nonumber\\
&  & {}  a^{\mu}_n~a_{\mu p} ~\sigma(m) ~T^i _{n,m}~ T^j_{-m,p} + a^{\mu}_n~a_{\mu o} ~\sigma(m) ~T^i _{n,m}~ T^j_{o,-m}] ~.
\end{eqnarray}

In equations (\ref{ext22}), a sum over repeated indices
(Latin indices $\neq 0$) and the identity $[AB,CD] = [A,C]DB + C[A,D]B + A[B,C]D + AC[B,D]$, were used.
Now if in the first term of the second expression we do $p \rightarrow n$, $m \rightarrow p$ and $n \rightarrow k$, in the second term $o \rightarrow n$, $m \rightarrow p$ and  $n \rightarrow k$ and finally in the fourth term
$o \rightarrow p$, we get:
\begin{eqnarray} \label{ext25}
\{\widehat{T}^i,\widehat{T}^j\}_{P.B} & = & \frac{i}{2}a^{\mu}_n ~a_{\mu p} \Big\{ \sigma(m) ~T^i _{n,m}~ T^j_{-m,p} + \sigma(m) ~T^i _{n,m}~ T^j_{p,-m}~+ {} \nonumber\\
& & {} \hspace{1.5cm}                                                \sigma(k) ~T^i _{k,p}~ T^j_{-k,n} + \sigma(k) ~T^i _{k,p}~ T^j_{n,-k} \Big\}~.
\end{eqnarray}
The operators in (\ref{ext23}) are manifestly symmetric for the
operators in (\ref{ext22}) are in fact commuting functions, thus
$\sqrt{2} \widehat{T}^i \equiv  a^{\mu}_n~T^i _{n,m}~a_{\mu m} ~ = a^{\mu}_m~T^i _{n,m}~a_{\mu n}$ and
relabeling the dummy indices $n \leftrightarrow m$ yields $T^i _{n,m} = T^i _{m,n}$.
Therefore if we do $k \leftrightarrow -m$ and promoting the Poisson brackets
to quantum mechanical commutators, then equation (\ref{ext25}) reads

\begin{eqnarray} \label{ext26}
[\widehat{T}^i,\widehat{T}^j] & = & \frac{1}{2}a^{\mu}_n ~a_{\mu p} \Big\{ \sigma(m) \big[ ~T^i _{n,m}~ T^j_{-m,p} - ~T^j _{n,m}~ T^i_{-m,p} \big]~+ {} \nonumber\\
& & {} \hspace{1.5cm} \sigma(m) \big[ ~T^i _{n,m}~ T^j_{-m,p} - ~T^j _{n,m}~ T^i_{-m,p} \big] \Big\} {} \nonumber\\
& = & a^{\mu}_n ~a_{\mu p} \Big\{ \sigma(m)  \big[~T^i _{n,m}~ T^j_{-m,p} - ~T^j _{n,m}~ T^i_{-m,p} \big]  \Big\}.
\end{eqnarray}

\subsection{A new matrix product}
Let us define the following matrix product.
\begin{equation} \label{star1}
(A \star B)_{np} \equiv  \sigma(k)~ A_{n,k}B_{-k,p}  ~,\end{equation}
which is perfectly associative and distributive
\begin{eqnarray}  \label{star2}
(A \star (B +C))_{i,j} & = &  \sigma(k)~A_{i,k}~(B_{-k,j} +C_{-k,j}) {} \nonumber\\
& = & {}  \sigma(k)~A_{i,k}~B_{-k,j} + \sigma(k)~A_{i,k}~C_{-k,j} \nonumber\\
& = & {} (A \star B)_{ij} + (A \star C)_{ij} ~,\end{eqnarray}
\begin{eqnarray}  \label{star3}
((A \star B) \star C)_{i,j} & = &  \sigma(m)~(A \star B)_{i,m}~C_{-m,j} {} \nonumber\\
& = & {}  \sigma(m)~ \sigma(k)~ A_{i,k}~B_{-k,m}~C_{-m,j} {} \nonumber\\
& = & {} \sigma(k)~A_{i,k}~( \sigma(m)~B_{-k.m}~C_{-m,j}) {} \nonumber\\
& = & {}  \sigma(k)~A_{i,k}~(B \star C)_{-k,j} = (A \star (B \star C))_{i,j} ~.
\end{eqnarray}
Now with this new matrix product $ \star$, it is easy to prove that:
\begin{equation} \label{ext27}
[\widehat{T}^i,\widehat{T}^j]^\star_{n,p} = ~\sum_{\mu} \sum_{n,p \neq 0} ~a^{\mu}_n ~[T^i, T^j]^\star_{n,p}~a_{\mu p} ~.\end{equation}
Therefore, with the matrix product introduced,
the operators $\widehat{T}^i$  are good ``generalized
isospin operators'' with respect to the $T_i$ ones, as
sketched in section \ref{sec:Isospin}. 
With this in hand we are ready to construct an algebra that resembles
the Virasoro Algebra (quadratic in creation/annihilation operators). 
All we have to do is to find matrices
(with components running from $-\infty$ to $+\infty$) that when multiplied via
the $\star$ product can be classified \emph{\`a la} Cartan.

\subsection{Connection with Virasoro algebra}

It is important to check that our generators include the Virasoro's.
To see this let's write Virasoro's generators in terms of the usual
creation/annihilation operators, $a^{\mu}_n$, according to
(\ref{alpha1}) and (\ref{alpha2}).
\begin{displaymath}
L_{m} = \left\{ \begin{array}{ll}
\frac{1}{2}  p^{2} + \frac{1}{2} \sum_{1}^{\infty} a^{\mu}_{-n}n~a^{\mu}_{n} \qquad & \textrm{for $m =0$,}
\\
\\
\frac{1}{2} \sum_{-\infty}^{\infty} a^{\mu}_{m-n}~ \sqrt{ |(m-n)n| }~a^{\mu}_{n} & \textrm{for $m \neq 0$.}
\end{array} \right.
\end{displaymath}
So comparing these with (\ref{ext23}) and choosing:
\begin{displaymath} \label{eqn:Viraconnection}
[T^m]_{ij} = \left\{ \begin{array}{ll}
\frac{1}{\sqrt{2}}~\{\delta_{i+0} \delta_{j+0} + i \delta_{i+j} \} \qquad & \textrm{for $m =0$,}
\\
\\
\frac{1}{\sqrt{2}}~ \sqrt{|ij|}~\delta_{i+j-m} & \textrm{for $m \neq 0$.}
\end{array} \right.
\end{displaymath}
we recover Virasoro generators.\\

\subsection{A suitable Lie algebra for this programme}
Let consider first a Lie algebra of three generators.
But before starting we have to study how it's matrix representations look like with our $\star$ product. First of all to keep in contact with the Virasoro algebra we want the generators that are to be identified with $L_0$ to be
diagonal, or more precisely hermitian and thus diagonalizable. So our diagonal
matrices are to be diagonal in the same sense as $L_0$ is, \ie
the entry $(0,0) \neq 0$ because from it we generate the dynamical term
$p^2$ so that to make sense out of $L_0$, the entries $(-n,n) \neq 0$ for
$n=1 \cdots \infty$ simply from looking at the expression for $L_0$ and finally
el the other entries equal to zero.
In the particular case of $SU(2)$ there is only one diagonal matrix belonging
to the Cartan subalgebra, $J_3$. So the basic elements of this algebra are:
\begin{displaymath}
J_3 =
\left( \begin{array}{ccc}
0 & 0 & \beta \\
0 & \gamma & 0 \\
\beta & 0 & 0
\end{array} \right)~,
\end{displaymath}
\begin{displaymath}
J_+ =
\left( \begin{array}{ccc}
r & s & t \\
s & u & v \\
t & v & w
\end{array} \right)~.
\end{displaymath}
The last matrix of the algebra, $J_-$, is obviously $J^\dag_+$. To these matrices we impose
$[J_3,J_\pm]^\star = \pm J_\pm$\footnote{It is interesting to see
that this is an eigenvalue problem in the adjoint representation of the
algebra but with the matrix product defined here. This needed be so because symmetrical matrices are closed
under conmutation for which our product $\star$ proved necessary.}
and $[J_+, J_-]^\star = J_3$. Which sets conditions upon the constants
involved.
Particularly for $J_3$ the conditions are $\gamma = 0$ or $\beta = 0$.
Let's remember that $J_3$ is to be associated with $L_0$ so necessarily
it must contain a dynamical term $p^2$, for it to be a proper candidate to the Hamiltonian. 
This term is to come
from the $(0,0)$ entry of the matrix so the solution demanding $\gamma = 0$ is
discarded. Now the second solution, $\beta = 0$, does not contain terms
proportional to the fields' excitations so this solution implies our
Hamiltonian predicts the dynamics of massless particles %
\footnote{ It is interesting
to note that the fact that our model will describe massless particles is
not only a consequence of it being a (BRST) gauge theory, rather, it is also
consequence of the choice above. That $\beta \neq 0$ would lead to a gauge
theory for massive HS fields is a novel feature though not clear yet and needs
to be explored in the future.}%
.%
This solution requires $-2sv = \gamma, \beta =0, u,r,t,w = 0$ and $s,v \neq 0$. So finally
we are left with a much simpler algebra than
$SU(2)$ namely $[J_3,J_+] = [J_3,J_-] = 0$ and $[J_+,J_-] = J_3$.
Now following our definition of the generators (\ref{ext23}) we can explicitly
write them in terms of the creation/annihilation operators of the fields'
excitations.
\begin{equation} \label{ext28}
\widehat{J}_3 =1/2 \gamma p^2 ~,\end{equation}
\begin{equation} \label{ext29}
\widehat{J}_+ = sp_\mu a^{\mu \dag}_1 ~,\end{equation}
\begin{equation} \label{ext30}
\widehat{J}_- = ep_\mu a^\mu_1 ~.\end{equation}
Demanding $J^\dag_+ = J_-$ then $s^\ast = e$,
($-2|s|^2 = \gamma$). That the algebra mentioned above is satisfied by these is
straightforward\footnote{This algebra is a proper Lie algebra that has been
studied already in ref. \cite{ham} p. 306}.

\section{Massless spin-1 and spin-2 fields} \label{sec:Massless spin1,2 fields}
The construction of the BRST charge using these three generator and
noting that two of the three commutators vanishes identically  yields:
\begin{equation} \label{ext31}
\mathcal{\widehat{Q}} = c_0 \widehat{J}_3 + c_{-1} \widehat{J}_- + c_1 \widehat{J}_+ - c_1 c_{-1} b_0~, \end{equation}
which is perfectly nilpotent. In this case we have no certainty about the
eventual degeneracy in ghost number of the vacuum state $|- \rangle $, but
we can be certain that it is physical, so analyzing $\mathcal{\widehat{Q}} |- \rangle = 0$,
we can conclude that $c_1 |- \rangle = b_1 |- \rangle = 0$,
$c_{-1} |- \rangle \neq 0 \neq b_{-1} |- \rangle$,
$c_0 |- \rangle = |+ \rangle$ and $b_0 |- \rangle = 0$.
Having considered that the vacuum is annihilated by $J_-$ y $J_3$.

Now the following step is to find the field equations induced by the BRST
symmetry of a certain scalar wave function $|A \rangle $, explicitly the symmetry
transformation induced is:
\begin{equation} \label{ext32}
|A \rangle \longrightarrow |A' \rangle = |A \rangle + \mathcal{\widehat{Q}} |\Lambda \rangle~, \end{equation}
where the transformation of the field $|A \rangle $ is
$\delta |A \rangle = \mathcal{\widehat{Q}} |\Lambda \rangle$. Obviously the field equations would be
given by the condition that $|A \rangle $ be physical \emph{i.e.},
$\mathcal{\widehat{Q}}|A \rangle=0 $.

A suitable gauge parameter $|\Lambda\rangle$ such that
$\mathcal{\widehat{Q}} |\Lambda \rangle = \delta |A \rangle$
has the correct ghost number is:
\begin{equation} \label{ext33}
|\Lambda \rangle = [b_{-1} \lambda(x) + b_{-1}a^{\mu \dag}_1 \Lambda_\mu(x)] |- \rangle ~.\end{equation}
From now on the subscript $1$ of the creation/annihilation operators from the
second term will be omitted, to this extent we will always be working to first level
in fields' excitations. Computing $\mathcal{\widehat{Q}} |\Lambda \rangle$ will give us an idea of
the field $|A \rangle $.
\begin{eqnarray} \label{ext35}
\mathcal{\widehat{Q}} |\Lambda \rangle & = &  c_0 \frac{\gamma}{2}p^2 b_{-1} \lambda(x) |- \rangle ~+~c_1 s p_\mu a^{\mu \dag} b_{-1} \lambda(x) |- \rangle {} \nonumber\\
& &  + \frac{\gamma}{2}p^2 \Lambda_\mu(x) a^{\mu \dag} c_0 b_{-1} | - \rangle + s^\ast p_\mu \Lambda_\mu (x) c_{-1}b_{-1} | - \rangle {} \nonumber\\
& & + \frac{s}{2}( p_\mu \Lambda_\nu (x)+ p_\nu \Lambda_\mu (x)) a^{\mu \dag}a^{\nu \dag} | - \rangle ~.\end{eqnarray}
Identifying amongst the terms above that appear multiplying all fields, those that are scalars, vectors, tensors, etc\dots, tells us that $|A \rangle $ should have
the following form:
\begin{eqnarray} \label{ext36}
|A \rangle & = &  \Omega(x) c_0b_{-1}|- \rangle ~ + ~ A_\mu (x) a^{\mu \dag} |- \rangle ~ + ~ \phi(x) |- \rangle {} \nonumber\\
& & +~\psi_\mu(x) a^{\mu \dag}_1 c_0 b_{-1}|- \rangle ~+ ~ \eta(x)c_{-1}b_{-1}|- \rangle{} \nonumber\\
& & +~h_{\mu \nu}(x) a^{\mu \dag}_1 a^{\nu \dag}_1 |- \rangle, \end{eqnarray}
such that the variations of the auxiliary fields
$\Omega(x)$, $A_\mu (x)$, $\phi(x)$, $\psi_\mu(x)$, $\eta(x)$, $h_{\mu \nu}(x)$
be:
\begin{equation} \label{ext37}
\delta \phi(x) =0~, \end{equation}
\begin{equation} \label{ext371}
\delta \Omega(x) = 1/2~\gamma p^2 \lambda(x)~, \end{equation}
\begin{equation} \label{ext372}
\delta A_\mu (x) = s p_\mu \lambda(x)~. \end{equation}
and
\begin{equation} \label{ext45}
\delta \psi_\nu (x) = \frac{\gamma p^2}{2}\Lambda_\nu(x)~, \end{equation}
\begin{equation} \label{ext46}
\delta \eta(x) = s^\ast p_\nu \Lambda_\nu (x)~, \end{equation}
\begin{equation} \label{ext47}
\delta h_{\mu \nu} (x) = \frac{s}{2}[p_\mu \Lambda_\nu(x) +p_\nu \Lambda_\mu(x) ]~. \end{equation}
As we said before the field equations are obtained imposing
$\mathcal{\widehat{Q}}|A\rangle=0$.
From equations (\ref{ext31}) and (\ref{ext36}) and demanding that
each of the different excitations of the vacuum state vanishes independently,
yields:
\begin{equation} \label{ext38}
\Box A_\mu(x) - \frac{2is}{\gamma} \partial_\mu \Omega(x) = 0~, \end{equation}
\begin{equation} \label{ext39}
\Box \phi(x) = 0~,\end{equation}
\begin{equation} \label{ext40}
\partial_\nu A_\nu (x) + \frac{i}{s^\ast} \Omega(x) = 0~,\end{equation}
\begin{equation} \label{ext48}
\Box \eta(x) - \frac{2is^\ast}{\gamma} \partial_\mu \psi_\mu(x) = 0~,\end{equation}
\begin{equation} \label{ext49}
\Box h_{\mu \nu}(x) - \frac{is}{\gamma}[\partial_\mu \psi_\nu (x) + \partial_\nu \psi_\mu(x)]= 0~, \end{equation}
\begin{equation} \label{ext50}
s~ \partial_\nu \eta(x) - 2s^\ast \partial_\mu h_{\mu \nu}(x) - i\psi_\nu (x) = 0~.\end{equation}
All of which are invariant under the transformations (\ref{ext37}) to
(\ref{ext47}). Now replacing (\ref{ext40}) in (\ref{ext38}) and since
$2|s|^2 = -\gamma$, we get:
\begin{equation} \label{ext41}
\partial_\nu (\partial_\mu A_\nu (x) - \partial_\nu A_\mu(x)) = \partial_\nu F_{\mu \nu} = F_{\mu \nu, \nu} = 0. \end{equation}
Thus we have recovered Maxwell's equations. Furthermore
solving for $h_{\mu \nu}(x)$ yields:
\begin{equation} \label{ext51}
\Box h_{\mu \nu} + h_{\lambda \lambda, \mu \nu} - (h_{\lambda \nu, \lambda \mu} + h_{\lambda \mu, \lambda \nu}) = 0~.\end{equation}
This last equation is nothing but Einstein's linearized equation of the
gravitational field where
$g_{\mu \nu} = \eta_{\mu \nu} + h_{\mu \nu}$ with
$ |h_{\mu \nu}| \ll 1$, which is very easy to see if we compare with
the Einstein's field equations for $h_{\mu \nu}$ (to first order in $h$) where
the matter term is present
\begin{equation} \label{ext52}
\Box h_{\mu \nu} + h_{\lambda \lambda, \mu \nu} - (h_{\lambda \nu, \lambda \mu} + h_{\lambda \mu, \lambda \nu}) = -16 \pi G S_{\mu \nu}~. \end{equation}

\section{Mixed symmetry $2^{nd}$ rank tensor field} \label{sec:Mixed 2nd rank}
Leaving aside the method used to construct the previous algebra and based on
its simplicity the generalization to an algebra with more generators is
almost immediate. However for the incorporation of mixed symmetry tensor
fields it suffices to consider a $5$-generator algebra
$\{\widehat{J}_{++}, \widehat{J}_+, \widehat{J}_0, \widehat{J}_-, \widehat{J}_{--} \}$ with $\widehat{J}_0 = \gamma p^2$,
$\widehat{J}_{++} = s p_\mu a^{\mu \dag}_2$ and so on\footnote{$[\widehat{J}_+, \widehat{J}_-]=\widehat{J}_0$ demands
$\gamma = -|s|^2$.}.
The BRST charge $\mathcal{\widehat{Q}}$ is constructed from the algebra (considering,
obviously that we have as many ghost field operators as generators in the algebra),
\begin{equation} \label{e00}
\mathcal{\widehat{Q}}=c_0\widehat{J}_0+c_1\widehat{J}_++c_2\widehat{J}_{++}+c_{-1}\widehat{J}_-+c_{-2}\widehat{J}_{--}-c_2c_{-2}b_0-c_1c_{-1}b_0
\end{equation}
and the procedure is essentially the one used previously, we start from the following gauge field:
\begin{equation} \label{e0}
|\Lambda \rangle = [b_{-2} \Lambda^{(1)}_\mu(x) a^{\mu \dag}_1 ~+~ b_{-1} \Lambda^{(2)}_\mu(x) a^{\mu \dag}_2]|- \rangle
\end{equation}
Then $\mathcal{\widehat{Q}} |\Lambda \rangle = \delta |A \rangle$ implies $|A \rangle$
should be:
\begin{eqnarray} \label{e1}
|A \rangle & = & [\psi^{(1)}_\mu(x) a^{\mu \dag}_1 c_0 b_{-2} + \psi^{(2)}_\mu(x) a^{\mu \dag}_2 c_0 b_{-1} {} \nonumber \\
& & {} +  M^{(12)}_{\mu \nu}(x) a^{\mu \dag}_1 a^{\mu \dag}_2 + \Omega^{(1)}(x)c_{-1} b_{-2} + \Omega^{(2)}(x)c_{-2} b_{-1}]|- \rangle
\end{eqnarray}
such that the variation of the fields are:
\begin{eqnarray} \label{e2}
\delta \psi^{(i)}_\mu(x) = - \gamma \Box \Lambda^{(i)}_\mu (x) {} \nonumber\\
\delta M^{(12)}_{\mu \nu}(x) = -is(\partial_\mu \Lambda^{(1)}_\nu(x)+ \partial_\nu \Lambda^{(2)}_\mu(x)) {} \nonumber \\
\delta \Omega^{(i)}(x) = -is^\ast \partial_\mu \Lambda^{(i)}_\mu(x) \end{eqnarray}
Note that $M^{(12)}_{\mu \nu}(x)$ does not have a definite symmetry.
Then demanding that $|A \rangle$ be physical, \emph{i.e.}
$\mathcal{\widehat{Q}}|A \rangle = 0$, we get:
\begin{equation} \label{e3}
-\gamma \Box M_{\mu \nu}(x) + i s \psi^{(2)}_{\mu, ~\nu}(x) + is \psi^{(1)}_{\nu, ~\mu}(x)=0
\end{equation}
\begin{equation} \label{e4}
-\gamma \Box \Omega^{(i)}(x) + is^\ast \psi^{(i) \mu}_{~~~,\mu}(x) = 0
\end{equation}
\begin{equation} \label{e5}
\psi^{(1/2)}_{~~~~\mu}(x) - is^{\ast} M^{~\sigma}_{\mu ~~, \sigma}(x) + is \Omega^{(2/1)}_{~~~~,\mu}(x) = 0
\end{equation}
Which simplify to:
\begin{equation} \label{e6}
M_{\mu \sigma, \nu \sigma}(x) + M_{\sigma \nu, \mu \sigma}(x) - \Box M_{\mu \nu}(x) - \Box^{-1} M_{\alpha \beta, \alpha \beta \nu \mu}(x) = 0
\end{equation}
\subsection{The  `axion' field}
If
$M_{\mu \nu}(x) = b_{\mu \nu}(x) + h_{\mu \nu}(x)$,
where $b_{\mu \nu} = 1/2(M_{\mu \nu} - M_{\nu \mu})$,
$h_{\mu \nu}= 1/2(M_{\mu \nu} + M_{\nu \mu})$,
then eqn. (\ref{e6}) read:
\begin{equation} \label{e7}
- H_{\mu \nu \sigma, \sigma}(x) - \Box h_{\mu \nu} - \Box^{-1} h_{\alpha \beta, \alpha \beta \mu \nu} +  h_{\mu \sigma, \nu \sigma} + h_{\sigma \nu, \mu \sigma} = 0.
\end{equation}
where we have made the following definition:
\begin{equation} \label{e8}
H_{\mu \nu \sigma}(x) \equiv b_{\mu \nu, \sigma}(x)+b_{\nu \sigma, \mu}(x)+b_{\sigma \mu,\nu}(x)
\end{equation}
Now, (\ref{e8}) has a purely symmetric and an antisymmetric part, therefore
each part must vanish separately:
\begin{equation} \label{e11}
H_{\mu \nu \sigma, \sigma}(x)=0
\end{equation}
\begin{equation} \label{e12}
- \Box h_{\mu \nu} - \Box^{-1} h_{\alpha \beta, \alpha \beta \mu \nu} +  h_{\mu \sigma, \nu \sigma} + h_{\sigma \nu, \mu \sigma} = 0
\end{equation}
Now, taking the trace of the last equation then:
\begin{equation} \label{e13}
\Box h_{\lambda \lambda} = h_{\alpha \beta, \alpha \beta}
\end{equation}
which if replaced in  the former equations yields:
\begin{equation} \label{e14}
-\Box h_{\mu \nu} -h_{\lambda \lambda, \mu \nu} +  h_{\mu \sigma, \nu \sigma} + h_{\sigma \nu, \mu \sigma}=0.
\end{equation}
Summarizing:\\
\ba
&H_{\mu \nu \sigma, \sigma}(x)=0\, ,&  \\
&-\Box h_{\mu \nu} -h_{\lambda \lambda, \mu \nu} +  h_{\mu \sigma, \nu \sigma} + h_{\sigma \nu, \mu \sigma}=0\,.&
\ea

\subsection{Gauge transformations of the physical fields}
From (\ref{e2}) we see that
$\delta M^{(12)}_{\mu \nu} = -is(\partial_\mu \Lambda^{(1)}_\nu + \partial_\nu \Lambda^{(2)}_\mu)$
then  we can know how the physical fields $b_{\mu \nu}, h_{\mu \nu}$
transform since $b_{\mu \nu}$ is the antisymmetric part of
$M^{(12)}_{\mu \nu}$ and $h_{\mu \nu}$ its symmetric part, then
\begin{eqnarray} \label{e18}
\delta b_{\mu \nu} & = & -\frac{is}{2} (\Lambda^{(1)}_{\nu ~,\mu}+\Lambda^{(2)}_{\mu ~,\nu}-\Lambda^{(1)}_{\mu ~,\nu}-\Lambda^{(2)}_{\nu ~,\mu}) {} \nonumber \\
& = & -\frac{is}{2} [(\Lambda^{(1)}_\nu - \Lambda^{(2)}_\nu)_{,\mu} - (\Lambda^{(1)}_\mu - \Lambda^{(2)}_\mu)_{,\nu}]  {} \nonumber \\
& = & -\frac{is}{2} (\varepsilon^{(-)}_{\nu ~~,\mu} - \varepsilon^{(-)}_{\mu ~~,\nu})
\end{eqnarray}
Similarly
\begin{equation} \label{e19}
\delta h_{\mu \nu}= -\frac{is}{2} (\varepsilon^{(+)}_{\nu ~~,\mu} + \varepsilon^{(+)}_{\mu ~~,\nu})
\end{equation}
% and
% \begin{eqnarray} \label{e20}
% \delta \varphi & = & \frac{1}{2d} \delta M_{\nu \nu} = \frac{is}{2d}(\Lambda^{(1)}_{\nu~,\nu}+ \Lambda^{(2)}_{\nu~,\nu}) {} \nonumber \\
% & = & {} \frac{is}{d}(\Lambda^{(1)}_\nu+ \Lambda^{(2)}_\nu)_{,\nu} = \frac{is}{d} \varepsilon^{(+)}_{\nu~~,\nu}
% \end{eqnarray}

Note that $h_{\munu}$ transforms the same
way the gravitational field under general coordinate transformation and
$b_{\munu}$ transforms just as the axion does,%
% See Polchinski VolI, pp29, Vol II pp. 76
% Also GSW I pp.215 (eqn 2.3.106) and eqns 3.4.46 - 3.4.57).
so the fact that
our equations were linearized versions of the field equations of these fields
was to be expected.

\section{HS gauge fields and the high-energy limit of string theory} \label{sec:higher rank}
Already in the previous section we could see that our first algebraic
construction was rather intricate and needed some refinement. To do this let
us write Virasoro generators with their dependence on the string constant
$\alpha'$ explicit:

\begin{displaymath}
L_{m} = \left\{ \begin{array}{ll}
\frac{1}{4}  \alpha' p^{2} + \frac{1}{2} \sum_{1}^{\infty} \alpha^{\mu}_{-n}\alpha_{\mu n} \qquad m =0,
\\
\\
\frac{1}{2} \{ \sqrt{2\alpha'} p_\mu \alpha^\mu_n + \sum_{-\infty}^{\infty} \alpha^{\mu}_{m-n}~ \alpha_{\mu n} \} \qquad m \neq 0 .
\end{array} \right.
\end{displaymath}
Now let us define the generators
\begin{equation} \label{e21}
\widehat{J}_0 \equiv \left[\lim_{\alpha' \rightarrow \infty} \frac{L_0}{\alpha'}=\frac{1}{4}p^2\right]
\end{equation}
\begin{equation} \label{e22}
\widehat{J}_m \equiv \left[\lim_{\alpha' \rightarrow \infty} \frac{L_m}{\sqrt{2\alpha'}}=\frac{1}{2}p_\mu\alpha^\mu_m\right]
\end{equation}
So now we have made an infinite-dimensional algebra, (whose truncation
provides the algebras with which we worked previously),
from the formal limit $\alpha' \to \infty$ of the rescaled 
Virasoro generators. From this construction, the generalization of the
above is immediate. Note that this rescaled and Virasoro-like
generators can be obtained as our original motivation, \ie, building
them like ``generalized isospin operators''. The form of these
would change and another formula like \ref{eqn:Viraconnection} would be obtained, whose 
specific expression is not relevant now.

Considering as gauge parameters the following:
\ba
\Lambda(x)&=&M_\mu a^{\mu \dag}_1\,, \nonumber\\ 
\Lambda^2(x)&=&M_\mu M_\nu a^{\mu \dag}_1 a^{\nu \dag}_1 = M_{(\mu \nu)} a^{\mu \dag}_1 a^{\nu \dag}_1\,,\nonumber\\
 \vdots \nonumber \\
\Lambda^n(x)&=&M_{\mu_1} \cdots M_{\mu_n} a^{\mu_1 \dag}_1 \cdots  a^{\mu_n \dag}_1 = M_{(\mu_1 \cdots \mu_n)} a^{\mu_1 \dag}_1 \cdots  a^{\mu_n \dag}_1\,.
\ea
and defining:
\begin{equation} \label{e23}
|\Lambda \rangle = b_{-1} e^{\Lambda(x)} |- \rangle\,,
\end{equation}
and following the BRST procedure, \emph{i.e.}
$\delta |A \rangle = \mathcal{\widehat{Q}}|\Lambda \rangle $
will provide a gauge invariant theory for symmetric tensor fields.
Obviously the previous results are included in this last procedure. 
For the case of higher spin and arbitrary symmetry this generalization
can also be accommodated as outlined in section \ref{sec:Mixed 2nd rank}.

As mentioned in the introduction, the connection between 
HS gauge fields and the high-energy (low-tension) limit
of string theory is a rather old idea. So is the BRST formulation
for the general theory of massless higher spin and arbitrary gauge fields.
However, the approach here
taken 
of building a BRST operator from ``generalized isospin-like operators''
is interesting in its own right. 

\section{Conclusions} \label{sec:Concl}
We have been able to construct a quantum field theory for massless
fields based upon the BRST symmetry induced by the 
``generalized isospin-like operators''. These operators
were built keeping close contact with certain properties of the Virasoro algebra,
which allowed us to find an extremely simple Lie algebra with which we could
obtain, without many complications, some very interesting results. 

Particularly
uncoupled fields $A_\mu$ and $h_{\mu \nu}$ corresponding to the photon and
graviton respectively. Besides the representations with which we worked
were symmetrical ones from which,
as a by-product we obtained a gauge invariant theory of symmetric tensor fields
through the nilpotent operator $\mathcal{\widehat{Q}}$ keeping close resemblance 
with the exterior
derivative operator which provides a gauge invariant theory but for
antisymmetric fields. The generalization to higher spins and arbitrary symmetry 
was also outlined.

Although our results (fields equations and transformation properties) are particular
cases of the  works on BRST formulation of massless higher spin gauge fields,
these are not our main results. Instead we want to stress that our approach is 
not only original but also makes contact with long established results, giving support
for our choice of ``generalized isospin-like operators'' as starting point and
thus raising the question of the future relevance in other contexts or applications
outside nuclear physics.

\section*{Acknowledgements}
The work of M.C. is partially supported by projects DGAPA-UNAM IN109107 and 
CONACYT \# 55310. He also 
wishes to thank DIPUC for partial support during the early stages of this work.
The work of JA  was
partially supported by Fondecyt \# 1060646.


\begin{thebibliography}{99}
%\cite{Yang:1954ek}
\bibitem{Yang:1954ek}
  C.~N.~Yang and R.~L.~Mills,
  %``Conservation of isotopic spin and isotopic gauge invariance,''
  Phys.\ Rev.\  {\bf 96} (1954) 191.
  %%CITATION = PHRVA,96,191;%%

%\cite{Heisenberg:1932dw}
\bibitem{Heisenberg:1932dw}
  W.~Heisenberg,
  %``On the structure of atomic nuclei,''
  Z.\ Phys.\  {\bf 77} (1932) 1.
  %%CITATION = ZEPYA,77,1;%%

%\cite{Fierz:1939ix}
\bibitem{Fierz:1939ix}
  M.~Fierz and W.~Pauli,
  %``On relativistic wave equations for particles of arbitrary spin in an
  %electromagnetic field,''
  Proc.\ Roy.\ Soc.\ Lond.\  A {\bf 173} (1939) 211.
  %%CITATION = PRSLA,A173,211;%%

%\cite{Fronsdal:1978rb}
\bibitem{Fronsdal:1978rb}
  C.~Fronsdal,
  %``Massless Fields With Integer Spin,''
  Phys.\ Rev.\  D {\bf 18} (1978) 3624.
  %%CITATION = PHRVA,D18,3624;%%

%\cite{Fang:1978wz}
\bibitem{Fang:1978wz}
  J.~Fang and C.~Fronsdal,
  %``Massless Fields With Half Integral Spin,''
  Phys.\ Rev.\  D {\bf 18} (1978) 3630.
  %%CITATION = PHRVA,D18,3630;%%

%\cite{Curtright:1980yk}
\bibitem{Curtright:1980yk}
  T.~Curtright,
  %``Generalized Gauge Fields,''
  Phys.\ Lett.\  B {\bf 165} (1985) 304.
  %%CITATION = PHLTA,B165,304;%%

%\cite{Ouvry:1986dv}
\bibitem{Ouvry:1986dv}
  S.~Ouvry and J.~Stern,
  %``Gauge Fields Of Any Spin And Symmetry,''
  Phys.\ Lett.\  B {\bf 177} (1986) 335.
  %%CITATION = PHLTA,B177,335;%%

%\cite{Koh:1986vg}
\bibitem{Koh:1986vg}
  I.~G.~Koh and S.~Ouvry,
  %``INTERACTING GAUGE FIELDS OF ANY SPIN AND SYMMETRY,''
  Phys.\ Lett.\  B {\bf 179} (1986) 115
  [Erratum-ibid.\  {\bf 183B} (1987) 434].
  %%CITATION = PHLTA,B179,115;%%

%\cite{Labastida:1986gy}
\bibitem{Labastida:1986gy}
  J.~M.~F.~Labastida and T.~R.~Morris,
  %``MASSLESS MIXED SYMMETRY BOSONIC FREE FIELDS,''
  Phys.\ Lett.\  B {\bf 180} (1986) 101.
  %%CITATION = PHLTA,B180,101;%%

%\cite{Labastida:1986ft}
\bibitem{Labastida:1986ft}
  J.~M.~F.~Labastida,
  %``Massless Bosonic Free Fields,''
  Phys.\ Rev.\ Lett.\  {\bf 58} (1987) 531.
  %%CITATION = PRLTA,58,531;%%

%\cite{Labastida:1987kw}
\bibitem{Labastida:1987kw}
  J.~M.~F.~Labastida,
  %``MASSLESS PARTICLES IN ARBITRARY REPRESENTATIONS OF THE LORENTZ GROUP,''
  Nucl.\ Phys.\  B {\bf 322} (1989) 185.
  %%CITATION = NUPHA,B322,185;%%

%\cite{Siegel:1985tw}
\bibitem{Siegel:1985tw}
  W.~Siegel and B.~Zwiebach,
  %``Gauge String Fields,''
  Nucl.\ Phys.\  B {\bf 263} (1986) 105.
  %%CITATION = NUPHA,B263,105;%%

%\cite{Witten:1985cc}
\bibitem{Witten:1985cc}
  E.~Witten,
  %``Noncommutative Geometry And String Field Theory,''
  Nucl.\ Phys.\  B {\bf 268} (1986) 253.
  %%CITATION = NUPHA,B268,253;%%

%\cite{Jensen:1996dc}
\bibitem{Jensen:1996dc}
  B.~Jensen and U.~Lindstrom,
  %``Classical interactions for tensionless strings,''
  Phys.\ Lett.\  B {\bf 398} (1997) 83
  [arXiv:hep-th/9612213].
  %%CITATION = PHLTA,B398,83;%%

%\cite{Francia:2002aa}
\bibitem{Francia:2002aa}
  D.~Francia and A.~Sagnotti,
  %``Free geometric equations for higher spins,''
  Phys.\ Lett.\  B {\bf 543} (2002) 303
  [arXiv:hep-th/0207002].
  %%CITATION = PHLTA,B543,303;%%

%\cite{Lindstrom:2003mg}
\bibitem{Lindstrom:2003mg}
  U.~Lindstrom and M.~Zabzine,
  %``Tensionless strings, WZW models at critical level and massless higher  spin
  %fields,''
  Phys.\ Lett.\  B {\bf 584} (2004) 178
  [arXiv:hep-th/0305098].
  %%CITATION = PHLTA,B584,178;%%

%\cite{Bonelli:2003kh}
\bibitem{Bonelli:2003kh}
  G.~Bonelli,
  %``On the tensionless limit of bosonic strings, infinite symmetries and
  %higher spins,''
  Nucl.\ Phys.\  B {\bf 669} (2003) 159
  [arXiv:hep-th/0305155].
  %%CITATION = NUPHA,B669,159;%%

%\cite{Sagnotti:2003qa}
\bibitem{Sagnotti:2003qa}
  A.~Sagnotti and M.~Tsulaia,
  %``On higher spins and the tensionless limit of string theory,''
  Nucl.\ Phys.\  B {\bf 682} (2004) 83
  [arXiv:hep-th/0311257].
  %%CITATION = NUPHA,B682,83;%%

%\cite{Pashnev:1997rm}
\bibitem{Pashnev:1997rm}
  A.~Pashnev and M.~M.~Tsulaia,
  %``Dimensional reduction and BRST approach to the description of a Regge
  %trajectory,''
  Mod.\ Phys.\ Lett.\  A {\bf 12} (1997) 861
  [arXiv:hep-th/9703010].
  %%CITATION = MPLAE,A12,861;%%

%\cite{Pashnev:1998ti}
\bibitem{Pashnev:1998ti}
  A.~Pashnev and M.~Tsulaia,
  %``Description of the higher massless irreducible integer spins in the  BRST
  %approach,''
  Mod.\ Phys.\ Lett.\  A {\bf 13} (1998) 1853
  [arXiv:hep-th/9803207].
  %%CITATION = MPLAE,A13,1853;%%

%\cite{Pashnev:1998sh}
\bibitem{Pashnev:1998sh}
  A.~I.~Pashnev and M.~M.~Tsulaia,
  %``On different BRST constructions for a given Lie algebra,''
  arXiv:hep-th/9810252.
  %%CITATION = HEP-TH/9810252;%%

%\cite{Buchbinder:2001bs}
\bibitem{Buchbinder:2001bs}
  I.~L.~Buchbinder, A.~Pashnev and M.~Tsulaia,
  %``Lagrangian formulation of the massless higher integer spin fields in  the
  %AdS background,''
  Phys.\ Lett.\  B {\bf 523} (2001) 338
  [arXiv:hep-th/0109067].
  %%CITATION = PHLTA,B523,338;%%

% %\cite{Francia:2002pt}
% \bibitem{Francia:2002pt}
%   D.~Francia and A.~Sagnotti,
%   %``On the geometry of higher-spin gauge fields,''
%   Class.\ Quant.\ Grav.\  {\bf 20} (2003) S473
%   [arXiv:hep-th/0212185].
%   %%CITATION = CQGRD,20,S473;%%
% 
% 
% 
% %\cite{Bekaert:2003az}
% \bibitem{Bekaert:2003az}
%   X.~Bekaert and N.~Boulanger,
%   %``On geometric equations and duality for free higher spins,''
%   Phys.\ Lett.\  B {\bf 561} (2003) 183
%   [arXiv:hep-th/0301243].
%   %%CITATION = PHLTA,B561,183;%%
% 
% 
% 
%\cite{Fotopoulos:2007nm}
\bibitem{Fotopoulos:2007nm}
  A.~Fotopoulos and M.~Tsulaia,
  %``Interacting Higher Spins and the High Energy Limit of the Bosonic String,''
  Phys.\ Rev.\  D {\bf 76} (2007) 025014
  [arXiv:0705.2939 [hep-th]].
  %%CITATION = PHRVA,D76,025014;%%

%\cite{Fotopoulos:2007yq}
\bibitem{Fotopoulos:2007yq}
  A.~Fotopoulos, N.~Irges, A.~C.~Petkou and M.~Tsulaia,
  %``Higher-Spin Gauge Fields Interacting with Scalars: The Lagrangian Cubic
  %Vertex,''
  JHEP {\bf 0710} (2007) 021
  [arXiv:0708.1399 [hep-th]].
  %%CITATION = JHEPA,0710,021;%%

%\cite{Bengtsson:2006pw}
\bibitem{Bengtsson:2006pw}
  A.~K.~H.~Bengtsson,
  %``Structure of Higher Spin Gauge Interactions,''
  J.\ Math.\ Phys.\  {\bf 48} (2007) 072302
  [arXiv:hep-th/0611067].
  %%CITATION = JMAPA,48,072302;%%

%\cite{Fotopoulos:2008ka}
\bibitem{Fotopoulos:2008ka}
  A.~Fotopoulos and M.~Tsulaia,
  %``Gauge Invariant Lagrangians for Free and Interacting Higher Spin Fields. A
  %Review of the BRST formulation,''
  arXiv:0805.1346 [hep-th].
  %%CITATION = ARXIV:0805.1346;%%
  
\bibitem{geo} H. Georgi: \emph{Lie Algebras in Particle Physics. From Isospin
to Unified Theories}. The Benjamin/Cummings Publishing Company, Inc. (1982).

\bibitem{ham} M. Hamermesh: \emph{Group Theory and its aplication to
physical problems}. Dover (1989).

\end{thebibliography}
\end{document}